\documentclass[preprint]{article}
\usepackage{graphicx}
\usepackage{dcolumn}
\usepackage{bm}
\usepackage{amsmath}
\usepackage{amssymb}
\usepackage{multirow}
\usepackage{caption}
\usepackage{epstopdf}
\usepackage{hyperref}
\usepackage{cite}

\begin{document}
{\setlength{\oddsidemargin}{1.2in}
\setlength{\evensidemargin}{1.2in} } \baselineskip 0.55cm
\begin{center}
{\LARGE {\bf Strange quark stars in mimetic gravitational theory }}
\end{center}
\date{\today}
\begin{center}
  Meghanil Sinha*, S. Surendra Singh \\
Department of Mathematics, National Institute of Technology Manipur,\\
Imphal-795004,India\\
Email:{ meghanil1729@gmail.com, ssuren.mu@gmail.com}\\
 \end{center}
 
\textbf{Abstract}: This paper presents a novel anisotropic quark star model in the backdrop of mimetic gravitational theory. Our study focuses on the strange quark stars using the MIT bag equation of state (EoS) admitting non-singular Buchdahl metric function. The proposed model matches the interior spacetime of the strange quark star with the exterior Schwarzschild spacetime. Our analysis of the energy conditions, radial and tangential EoSs along with the energy momentum tensor gradients and TOV equilibrium condition analysis support the model's physical validity. Further study of adiabatic index, surface redshift function and speed of sound analysis have demonstrated the stability of the strange quark star in mimetic gravity. Thus, we can say that the model stability and consistency have been validated for various parameter values of the Buchdahl function throughout our study in the context of mimetic gravity.\\


\section{Introduction}\label{sec1}

\hspace{0.8cm} New findings propose potential limitations to Einstein's general relativity (GR), sparking modification discussions. High-gravity phenomena and renormalization issues highlight potential GR limitations. GR falls short in accounting for the Universe's accelerating expansion. GR accommodates accelerated expansion with the dark matter (DM) and dark energy (DE) hypotheses. Although experimental evidence for DM and DE remains elusive. Modifying Einstein's GR preserves its successes while addressing limitations. Despite challenges, GR's successes in the solar system is still well established. GR's prominence is reinforced by the recent breakthroughs in the black hole imaging and gravitational waves from the M-87 Event Horizon Telescope \cite{K1,K2}. In spite of this GR still has limitations and unexplained aspects.\\
Inflation theory provides compelling solutions to cosmic challenges like flatness, horizon and fine-tuning problems \cite{A1,A2,A3}. Inflation provides a framework for understanding cosmic microwave background's large scale structure formations from initial fluctuations \cite{A4,A6,A7}. A scalar field facilitates inflation in GR theory \cite{A4}. Studies have validated scalar field model's alignment with empirical evidence \cite{A8,A9}. But GR still encounters difficulties in reconciling with accelerated expansion observations \cite{A12,A13,A14}. Modified gravitational theories offer explanations for the galaxy rotation curves and cosmic acceleration without invoking DM or DE \cite{A15,A18,A20,A22}. Modified gravity theories introduce scale-dependent effects preserving GR's success at solar system scales. The mimetic DM model offers an alternative perspective on DM's role in cosmology \cite{A23}. The model isolates the scalar degree of freedom from the metric forming its foundation. The model provides solutions for DM, DE and singularity resolution with a ghost-free massive gravity formulation \cite{A24,A25,A26}. This mimetic gravity theory, a Weyl-symmetric GR extension, enables a perfect fluid to emulate cold DM \cite{A23}. Geometric insights from this mimetic gravity shed light on late-time acceleration, inflation and DM \cite{N63,N64}. In gravitating systems, mimetic scalar field adds a longitudinal degree of freedom to standard transverse models. The longitudinal mode plays the role of mimetic DM. It isolates the gravity's conformal degree of freedom using the physical metric (auxiliary metric) and with the help of the scalar field \cite{A23}. The conformal degree of freedom's dynamical nature persists in the absence of matter yielding non-trivial solutions \cite{A23}. Mimetic gravity's unique features have led to increase attention and research advancements in recent times. From a cosmological perspective, this gravity provides a robust framework for understanding flat galaxy rotation curves, cosmological singularities along with gravitational waves \cite{A25,N65,N67}. This theory addresses black hole singularities and introduces new gravitational solutions \cite{A24,PS}. Thus, exploring mimetic gravity's astrophysical consequences has now become increasingly crucial. Upcoming detections may reveal non-standard signatures deviating from GR. Given these motivations, we examine the theory's astrophysical effects in this manuscript. Compact objects, characterized by extreme density and gravity enable testing ground for these alternative theories of gravity \cite{Q33,Q34,Q35,MP}. Study of non-singular compact objects constrains relativistic EoS in GR and also within modified gravity via the energy conditions and stability analysis \cite{IA,AP,RP}. Stability assessment employs adiabatic index, surface redshift and speed of sound analysis. Our analysis centers on a particular class of compact stars namely the strange quark stars with the MIT bag EoS.\\
Emerging evidence from pulsars, X-ray bursts and NICER points to exotic matter in compact stars \cite{D31,D32}. Early speculation suggested that strange quark matter would comprise part or all of the compact object \cite{D34,D35}. This type of stars are referred to as strange stars. Strange quark matter comprises up(u), down(d) and strange(s) quarks in roughly equal proportions with electrons balancing the charge. The MIT bag model effectively describes the strange quark matter \cite{D36}. In this configuration, the bag is a spherically symmetric space where the quarks are confined. Strongly interacting quarks favor a three-flavor system (u, d and s) over a two flavor system for stability. Self-bound stability in quark stars necessitates a considerable proportion of strange quarks. Researchers have investigated strange quark stars in Einstein gravity and beyond, in the modified theories of gravity utilizing the MIT bag model \cite{D37,D39,D42,D45,D47,D,M,Q,B}. Quark stars enable investigation of deconfinement phase transitions in neutron star interiors. In the light of this background, this paper offers a unique strange quark star solution in the framework of mimetic gravitational theory.\\
This paper is organized in the following manner: We have provided an overview of the modified gravitational theories along with a detailed discussion on the mimetic gravity in section (\ref{sec1}). In this section our discussion also includes non-singular compact strange stars as well. Section (\ref{sec2}) covers the application of mimetic gravity's field equations to spherically symmetric spacetimes. Assuming Buchdahl symmetry, we have matched the interior with the exterior Schwarzschild spacetime to derive the Buchdahl metric coefficients. The MIT bag EoS is discussed in section (\ref{sec3}) for our strange quark star model. In section (\ref{sec4}) we have analyzed the energy conditions, EoSs for radial and tangential pressures and equilibrium conditions via TOV equations. Section (\ref{sec5}) explores the stability through adiabatic index, surface redshift function and via speed of sound analysis. The main conclusion and highlights of our research are discussed in section (\ref{sec6}).\\ 

\section{Mimetic gravity formalism}\label{sec2}

\hspace{0.8cm}Mimetic DM emerged in the research works of \cite{A27,A28,A29}. Here we have the physical metric $ g_{\xi\upsilon} $ in terms of the $ \bar{g}_{\xi\upsilon} $ (auxiliary metric) and the scalar field $ \eta $ as \cite{A23}
\begin{equation}\label{1}
g_{\xi\upsilon} = -(\bar{g}_{\alpha\beta}\partial^{\alpha}\eta \partial^{\beta}\eta)\bar{g}_{\xi\upsilon} 
\end{equation}\\
which ensures that the conformal transformations of the auxiliary metric leave the physical metric unchanged. Consistency requirements lead to a constraint on the mimetic scalar field as\\
\begin{equation}\label{2}
g^{\xi\upsilon} \partial_{\xi}\eta \partial_{\upsilon}\eta = -1.
\end{equation}\\
The mimetic theory's four dimensional action is expressed as\\
\begin{equation}\label{3}
\emph{S} = \frac{1}{2\kappa} \int d^{4}x\sqrt{-g(\bar{g}_{\alpha\beta},\eta)}\emph{R}(\bar{g}_{\alpha\beta},\eta) + \int d^{4}x \sqrt{-g(g_{\alpha\beta},\eta)} L_{m}
\end{equation}\\
where $ \kappa = \frac{8 \pi G}{c^{4}} $, $ G = $ Newtonian constant, $ c= $ speed of light, $ g(\bar{g}_{\alpha\beta},\eta) = $ the determinant of the metric tensor, $ \eta = $ scalar field, $ \emph{R} $ = Ricci scalar, and $ L_{m}= $ matter lagrangian. Equation (\ref{3}) yields the field equations for this theory as\\
\begin{eqnarray}\label{4}
G_{\alpha}^{\beta} - (G-T)\partial_{\alpha}\eta \partial^{\beta}\eta = \kappa T_{\alpha}^{\beta}
\nonumber \\
\nabla_{\alpha} [(G-T)\partial^{\alpha}\eta] = 0
\end{eqnarray}\\
where $ G_{\alpha\beta} = $ Einstein tensor, $ T_{\alpha\beta} = $ stress energy tensor and $ G $ and $ T $ indicate the traces of these components respectively. The energy-momentum tensor for the anisotropic configurations is represented by\\
\begin{equation}\label{5}
T_{\alpha}^{\beta} = (p_{t} + \rho)u_{\alpha}u^{\beta} + p_{t}\delta_{\alpha}^{\beta} + (p_{r} - p_{t})\zeta_{\alpha}\zeta^{\beta}
\end{equation}\\
with $ u_{\alpha} $ as the timelike vector with $ u_{\alpha} = [1, 0, 0, 0] $ and $ \zeta_{\alpha} $ represents the unit radial vector (spacelike) with $ \zeta^{\alpha} = [0, 1, 0, 0] $, which satisfies $ u^{\alpha}u_{\alpha} = -1 $ and $ \zeta^{\alpha}\zeta_{\alpha} = 1 $. Our objective is to utilize the field equations in spherically symmetric spacetime given by
\begin{equation}\label{6}
ds^{2} = e^{i(r)}dt^{2} - e^{j(r)}dr^{2} - r^{2}(d \theta^{2} + \sin^{2} \theta d \phi^{2})
\end{equation}\\
where $ e^{i(r)} $ and $ e^{j(r)} $ represent undetermined functions. Field equations (\ref{4}) applied to the spactime equation (\ref{6}) gives the set of field equations as\\
\begin{equation}\label{7}
\frac{e^{-j}}{r^{2}}(-1 + e^{j} + j'r) = 8 \pi \rho
\end{equation}
\begin{eqnarray}\label{8}
\frac{e^{-j}i'}{r} - \frac{1}{r^{2}} + \frac{e^{-j}}{r^{2}} + \eta'^{2}\Big( 8 \pi \rho e^{-j} - 8 \pi p_{r}e^{-j} - 16 \pi p_{t} e^{-j} - \frac{2(e^{-j})^{2}}{r} - \frac{2e^{-j}}{r^{2}} + \frac{2(e^{-j})^{2}}{r^{2}}
\nonumber\\
\frac{2i'(e^{-j})^{2}}{r^{2}} - \frac{(e^{-j})^{2}i'j'}{2} + (e^{-j})^{2}i'' + e^{i}(e^{-j})^{2}(i')^{2} - \frac{(e^{-j})^{2}(i')^{2}}{2} \Big) = -8 \pi p_{r}
\end{eqnarray}
\begin{equation}\label{9}
\frac{e^{-j}j'}{2r} - \frac{e^{-j}i'}{2r} + \frac{e^{-j}i'j'}{4} - \frac{e^{-j}i''}{2} + \frac{3e^{-j}(i')^{2}}{4} = -8 \pi p_{t}
\end{equation}\\ 
where $ ' $ represents the derivative w.r.t $ r $. Thus we have three equations in six unknown functions $ e^{i(r)}, e^{j(r)}, \rho, p_{r}, p_{t} $ and $ \eta $. Numerical calculations require supplementary conditions. A specific form for the metric potential $ e^{i(r)} $ has been adopted in our work. Our chosen special form follows the Buchdahl metric, analogous to \cite{B,B21}
\begin{equation}\label{10}
e^{i(r)} = \frac{\Sigma(\Xi r^{2} + 1)}{\Sigma + \Xi r^{2}}, \hspace{2cm} 0 < \Sigma < 1
\end{equation}
where $ \Xi $ has a free parameter of the metric function. The metric function and its derivative are well-behaved and non-singular at the stellar center as \\
\begin{eqnarray}\label{11}
e^{i(0)} = 1
\nonumber \\
\partial_{r}e^{i(0)}\big|_{r=0} = 0.
\end{eqnarray}
Junction conditions at $ r= R $, where the interior spacetime meets the exterior Schwarzschild spacetime allows us to find $ \Xi $, where $ M $ denotes the total stellar mass. The exterior Schwarzschild spacetime is given as
\begin{equation}\label{12}
ds^{2} = \Big(1- \frac{2M}{r}\Big)dt^{2} - \frac{1}{\Big(1- \frac{2M}{r}\Big)}dr^{2} - r^{2}\Big(d\theta^{2} + \sin ^{\theta}d\phi^{2}\Big).
\end{equation}
From the continuity of the $ g_{rr} $ component, we have
\begin{equation}\label{13}
\big(1- \frac{2M}{R}\Big)^{-1} = \frac{\Sigma(\Xi R^{2} + 1)}{\Sigma + \Xi R^{2}}
\end{equation}
and from the continuity of $ \frac{\partial g_{rr}}{dr} $, we have
\begin{equation}\label{14}
\frac{-2M}{(R-2M)^{2}} = \frac{2\Xi(\Sigma - 1)\Sigma R}{(\Xi R^{2} + \Sigma)^{2}}.
\end{equation}
From the above equations, the solution for $ \Xi $ is provided as 
\begin{equation}\label{15}
\Xi = -\frac{2\Sigma M}{R^{2}(2\Sigma M - \Sigma R + R)}.
\end{equation}
We could observe that $ \Xi $ is negative for realistic stellar mass and radius combinations. Figure (\ref{1}) displays the matched metric potential across interior and exterior regions. With small positive $ \Sigma $(the free parameter), the Buchdahl metric potential becomes physically viable. A black vertical line indicates the junction point here. Now we choose a scalar field form which would satisfy equation (\ref{2}), that is given by
\begin{equation}\label{16}
\eta = \int \frac{1}{\sqrt{e^{i}}} dr. 
\end{equation}
With the help of equation (\ref{16}) and equation (\ref{10}) we can approach for the numerical solutions for the set of equations (\ref{7}-\ref{9}).
\begin{figure}[ht!]
\centering
\includegraphics[scale=0.5]{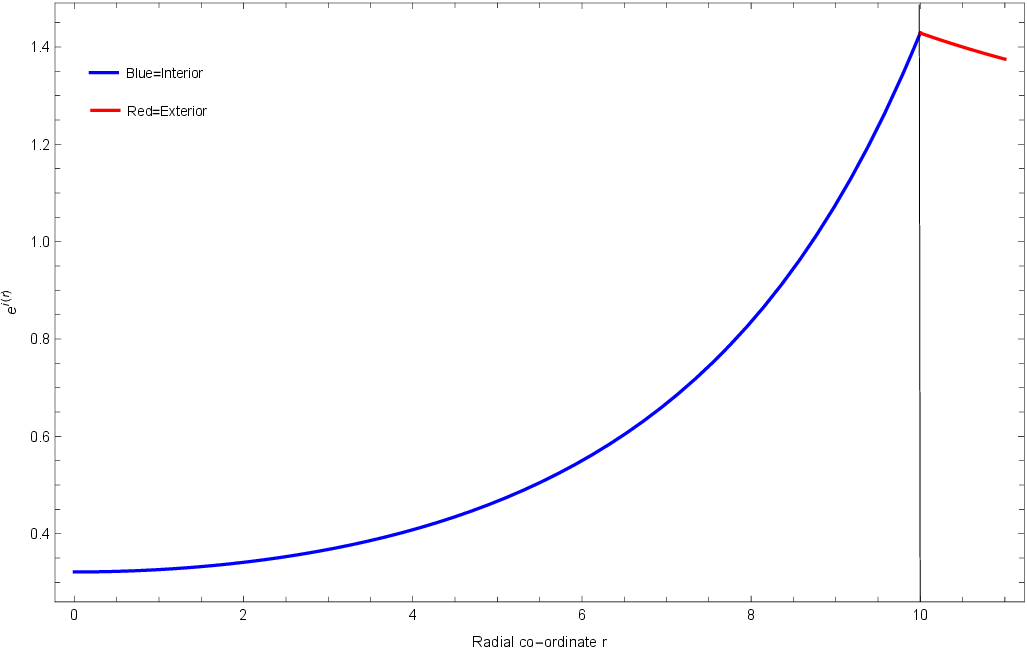}
\caption{Figure of the matched interior spacetime with the exterior spacetime w.r.t radius for $ PSR J1614-2230 $ with $ \Sigma = 0.0000135 $}\label{1}
\end{figure}\\

\section{MIT bag EoS}\label{sec3}

\hspace{0.8cm}Specifying sources are a prerequisite for proceeding with the structure equations. The MIT bag model characterizes quark matter inside the strange stars \cite{Q55}. The bag concept represents a finite volume where asymptotically free quarks are confined. The MIT bag constant has observationally imposed bounds within $[41 MeV/fm^{3}, 75 MeV/fm^{3}]$ \cite{B22,B23}. The bag constant $ \texttt{B} $ provides the inward pressure necessary to confine quarks inside the bag. The MIT bag model assumes massless, non-interacting u, d and s quarks. The MIT bag model defines the radial pressure as\\
\begin{equation}\label{17}
p_{r} = \sum_{h=u,d,s} p^{h} - \texttt{B}
\end{equation}
and energy density as
\begin{equation}\label{18}
\rho = \sum_{h=u,d,s} \rho^{h} + \texttt{B}
\end{equation}
where $ p^{h} $ and $ \rho^{h} $ denote the radial pressure and energy density per quark flavor respectively. The simplified MIT bag EoS model is derived by combining the above equations as\\
\begin{equation}\label{19}
p_{r} = \frac{1}{3} (\rho - 4 \texttt{B}).
\end{equation}\\
In line with the previous works, we set $ \texttt{B} = 0.00001052 km^{-2} $ \cite{B,B24}. With equation (\ref{19}) in hand, scalar field from equation (\ref{16}) and Buchdahl metric function from equation (\ref{10}), we can numerically study the compact stellar solutions with the help of the graphical representations. Field equations (\ref{7}) and (\ref{8}) along with the initial condition $ e^{j} \big|_{r=R} = e^{-i} \big|_{r=R} $ yield the metric function which is shown in figure (\ref{2}) for different parameter values. Physically valid behavior is observed for the metric potential throughout.\\
\begin{figure}[ht!]
\centering
\includegraphics[scale=0.5]{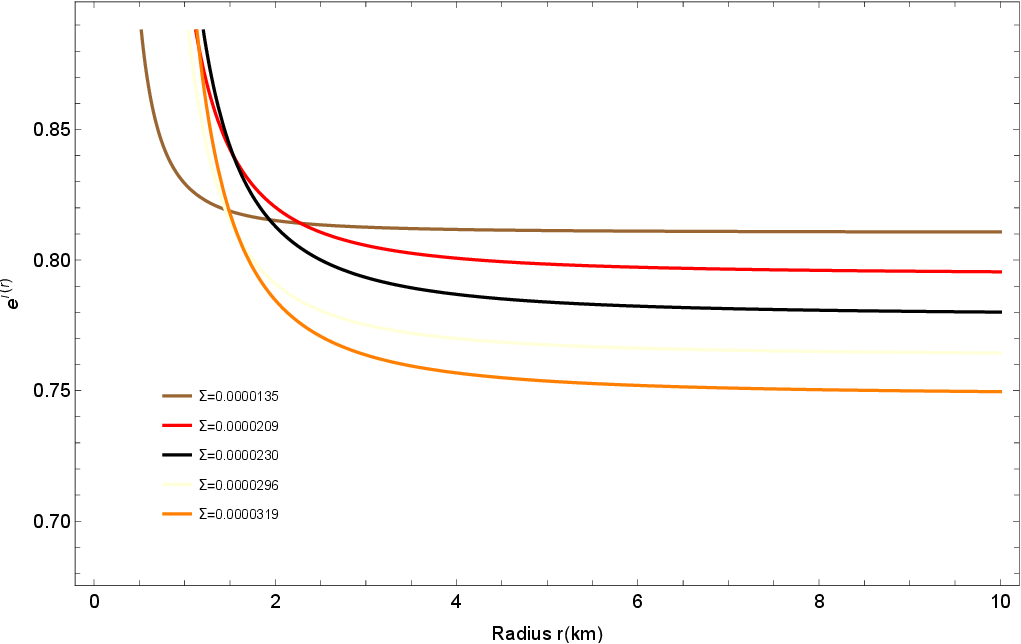}
\caption{Graph of the metric potential for $ PSR J1614-2230 $ with $ \Sigma = 0.0000135, \Xi = -5.84604 \times 10^{-8} $, $ 4U 1608-52 $ with $ \Sigma = 0.0000209, \Xi = -9.04237 \times 10^{-8} $, $ SMC X-1 $ with $ \Sigma = 0.0000230, \Xi = -8.36387 \times 10^{-8} $, $ 4U 1538-52 $ with $ \Sigma = 0.0000296, \Xi = -1.05825 \times 10^{-8} $, $ SAX JI8084.4-3658 $ with $ \Sigma = 0.0000319, \Xi = -7.66959 \times 10^{-8} $ }\label{2}
\end{figure}\\

\section{Different physical features of the strange quark star model in mimetic theory}\label{sec4}

\hspace{0.8cm}In this section we have explored various physical aspects which characterizes our stellar model such as energy conditions, radial and transversal equation of states and energy momentum tensor gradients.\\

\subsection{Energy conditions}

\hspace{0.5cm}Energy conditions assess the realism of matter behavior. There are mainly four types of energy conditions namely:\\

\begin{enumerate}
\item \textbf{Null energy condition} (NEC)
\item \textbf{Weak energy condition} (WEC)
\item \textbf{Strong energy condition} (SEC)
\item \textbf{Dominant energy condition} (DEC)
\end{enumerate}
Validation of NEC everywhere is crucial, as it is a basic requirement for WEC and SEC \cite{B25,B26}. The energy conditions are formulated using the energy density and anisotropic pressure as\\
\begin{itemize}
\item \textbf{NEC}: $ \rho + p_{r} \geq 0 $ and $ \rho + p_{t} \geq 0 $
\item \textbf{WEC}: $ \rho > 0 $ and $ \rho + p_{r} \geq 0 $ and $ \rho + p_{t} \geq 0 $
\item \textbf{DEC}: $ \rho - \mid p_{r} \mid \geq 0 $ and $ \rho - \mid p_{t} \mid \geq 0$
\item \textbf{SEC}: $ \rho + p_{r} + 2 p_{t} \geq 0 $
\end{itemize} 
Figures (\ref{3}-\ref{7}) illustrate the energy conditions for the strange quark star candidates $ PSR J1614-2230 $, $ 4U 1608-52 $, $ SMC X-1 $, $ 4U 1538-52 $ and $ SAX JI8084.4-3658 $ in mimetic gravity. Clearly we can see that each of the energy conditions are fulfilled within the stellar interior across different parameters which indicate that our stellar model is physically plausible in this scenario consisting of strange quarks with the MIT bag EoS and Buchdahl metric. 
\begin{figure}[ht!]
\centering
\includegraphics[scale=0.5]{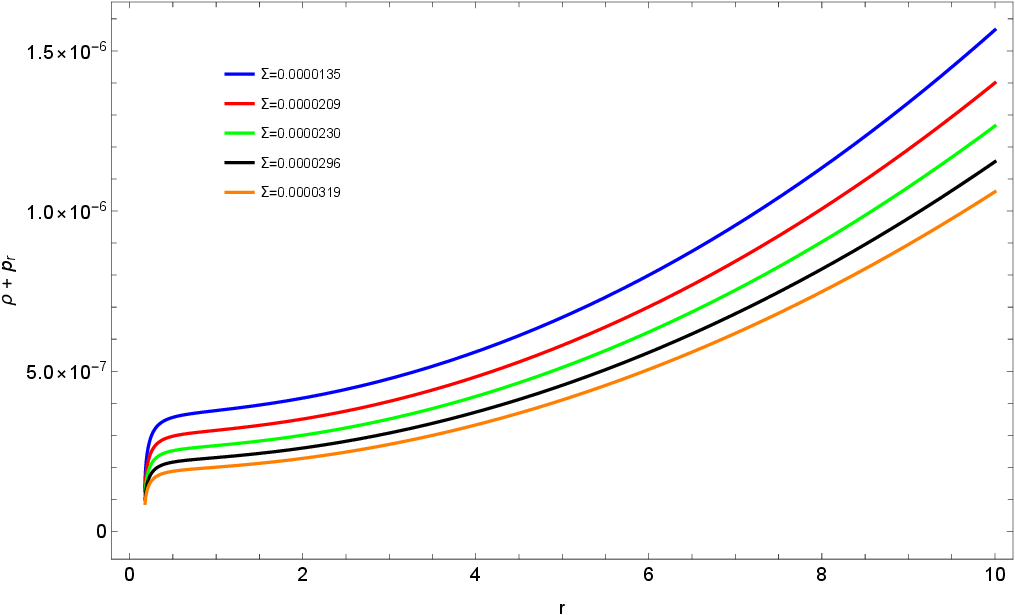}
\caption{Graph of $ \rho + p_{r} $ against $ r (km) $ for $ \Sigma = 0.0000135, \Xi = -5.84604 \times 10^{-8} $, $ \Sigma = 0.0000209, \Xi = -9.04237 \times 10^{-8} $, $ \Sigma = 0.0000230, \Xi = -8.36387 \times 10^{-8} $, $ \Sigma = 0.0000296, \Xi = -1.05825 \times 10^{-8} $, $ \Sigma = 0.0000319, \Xi = -7.66959 \times 10^{-8} $}\label{3}
\end{figure}\\
\begin{figure}[ht!]
\centering
\includegraphics[scale=0.5]{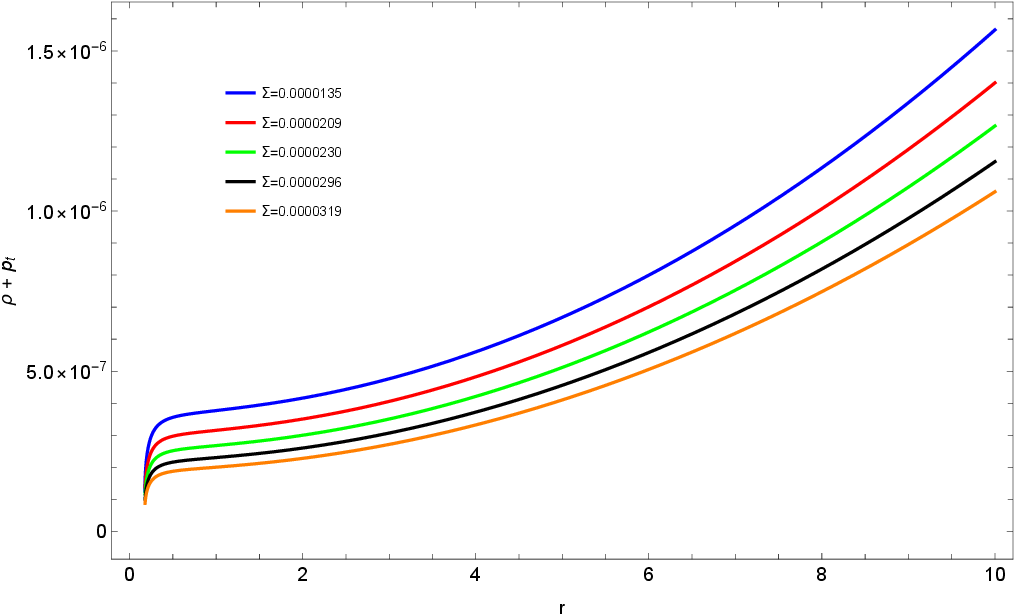}
\caption{Graph of $ \rho + p_{t} $ against $ r (km) $ for $ \Sigma = 0.0000135, \Xi = -5.84604 \times 10^{-8} $, $ \Sigma = 0.0000209, \Xi = -9.04237 \times 10^{-8} $, $ \Sigma = 0.0000230, \Xi = -8.36387 \times 10^{-8} $, $ \Sigma = 0.0000296, \Xi = -1.05825 \times 10^{-8} $, $ \Sigma = 0.0000319, \Xi = -7.66959 \times 10^{-8} $}\label{4}
\end{figure}\\
\begin{figure}[ht!]
\centering
\includegraphics[scale=0.5]{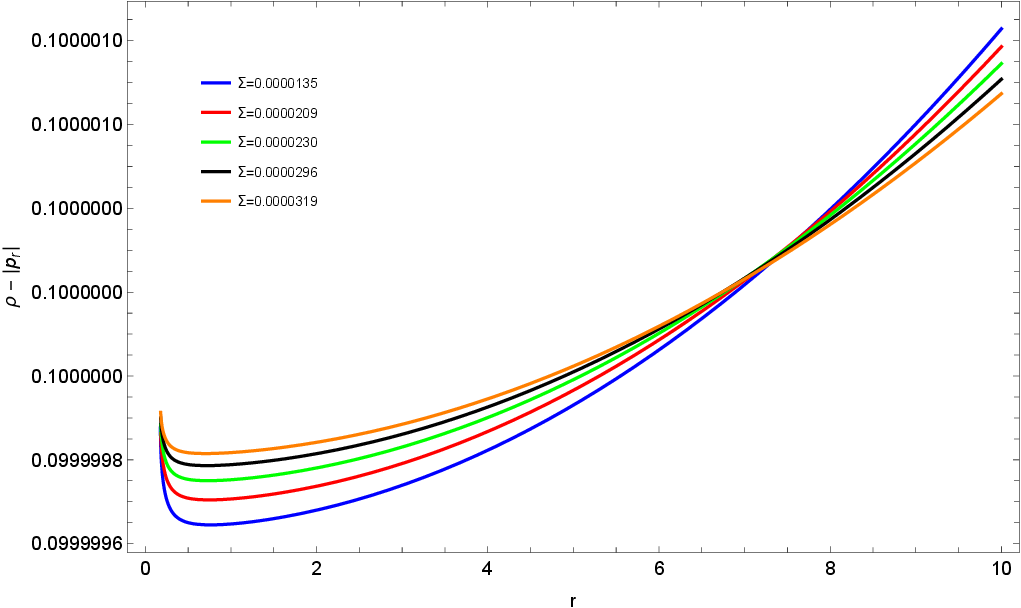}
\caption{Graph of $ \rho - \mid p_{r} $ against $ r (km) $ for $ \Sigma = 0.0000135, \Xi = -5.84604 \times 10^{-8} $, $ \Sigma = 0.0000209, \Xi = -9.04237 \times 10^{-8} $, $ \Sigma = 0.0000230, \Xi = -8.36387 \times 10^{-8} $, $ \Sigma = 0.0000296, \Xi = -1.05825 \times 10^{-8} $, $ \Sigma = 0.0000319, \Xi = -7.66959 \times 10^{-8} $}\label{5}
\end{figure}\\
\begin{figure}[ht!]
\centering
\includegraphics[scale=0.5]{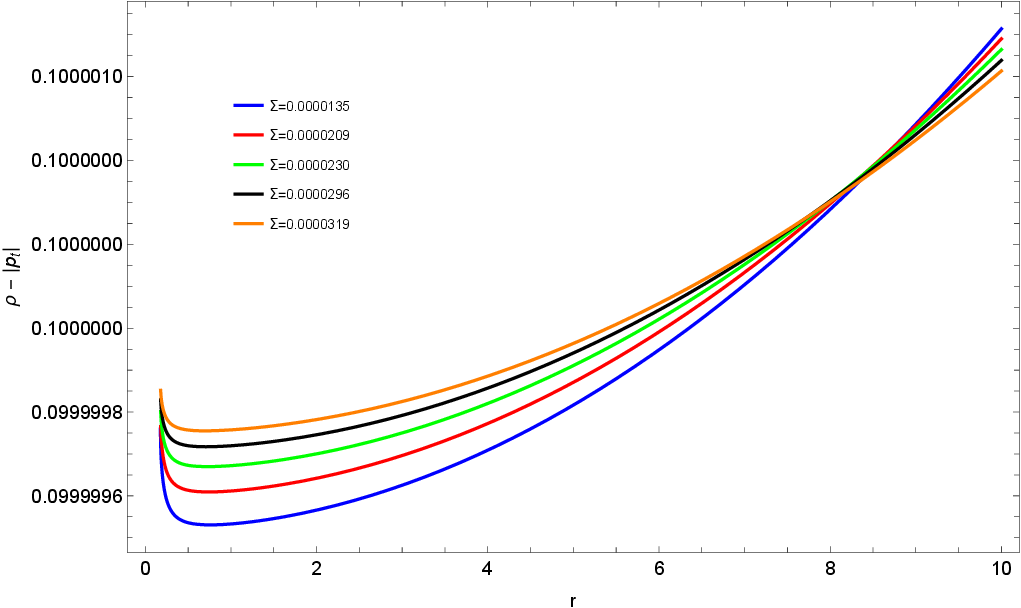}
\caption{Graph of $ \rho - \mid p_{t} $ against $ r (km) $ for $ \Sigma = 0.0000135, \Xi = -5.84604 \times 10^{-8} $, $ \Sigma = 0.0000209, \Xi = -9.04237 \times 10^{-8} $, $ \Sigma = 0.0000230, \Xi = -8.36387 \times 10^{-8} $, $ \Sigma = 0.0000296, \Xi = -1.05825 \times 10^{-8} $, $ \Sigma = 0.0000319, \Xi = -7.66959 \times 10^{-8} $}\label{6}
\end{figure}\\
\begin{figure}[ht!]
\centering
\includegraphics[scale=0.5]{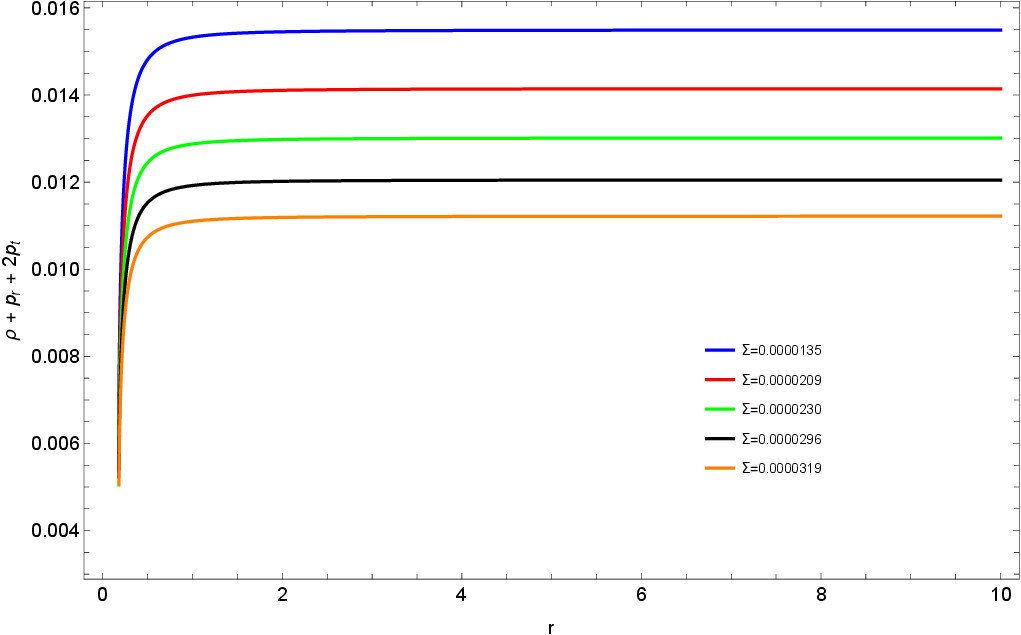}
\caption{Graph of $ \rho + p_{r} + 2 p_{t} $ against $ r (km) $ for $ \Sigma = 0.0000135, \Xi = -5.84604 \times 10^{-8} $, $ \Sigma = 0.0000209, \Xi = -9.04237 \times 10^{-8} $, $ \Sigma = 0.0000230, \Xi = -8.36387 \times 10^{-8} $, $ \Sigma = 0.0000296, \Xi = -1.05825 \times 10^{-8} $, $ \Sigma = 0.0000319, \Xi = -7.66959 \times 10^{-8} $}\label{7}
\end{figure}\\ 

\subsection{Equations of state}

\hspace{0.5cm}Next we focus on the EoS parameter $ \varpi $, along with energy density gradients and anisotropic pressure gradients. The anisotropic EoS parameter is formulated below as
\begin{eqnarray}\label{20}
\varpi_{r} = \frac{p_{r}}{\rho}
\nonumber \\
\varpi_{t} = \frac{p_{t}}{\rho}.
\end{eqnarray}
The numerical solutions for these equations are visualized in the figures (\ref{8}) and (\ref{9}) respectively. The EoS parameters fall within the viable range of $ 0 $ to $ 1 $, indicating a physically plausible cosmological fluid. The EoS parameter signifies dust nature at $ 0 $ and Zel'dovich stiff fluid at $ 1 $. And then figures (\ref{10}-\ref{12}) show the variation of the radial gradients of energy momentum tensor components. The gradient's positive behavior and the absence of singularities with a smooth nature confirm a viable stellar structure.\\
\begin{figure}[ht!]
\centering
\includegraphics[scale=0.5]{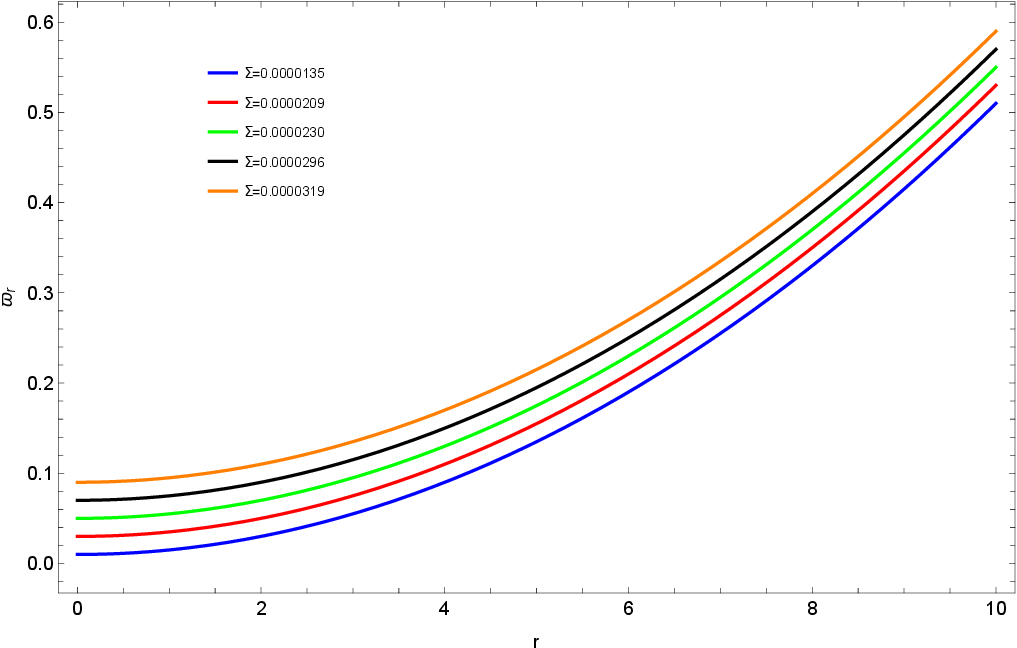}
\caption{Variation of $ \varpi_{r} $ for different parameter values}\label{8}
\end{figure}\\
\begin{figure}[ht!]
\centering
\includegraphics[scale=0.5]{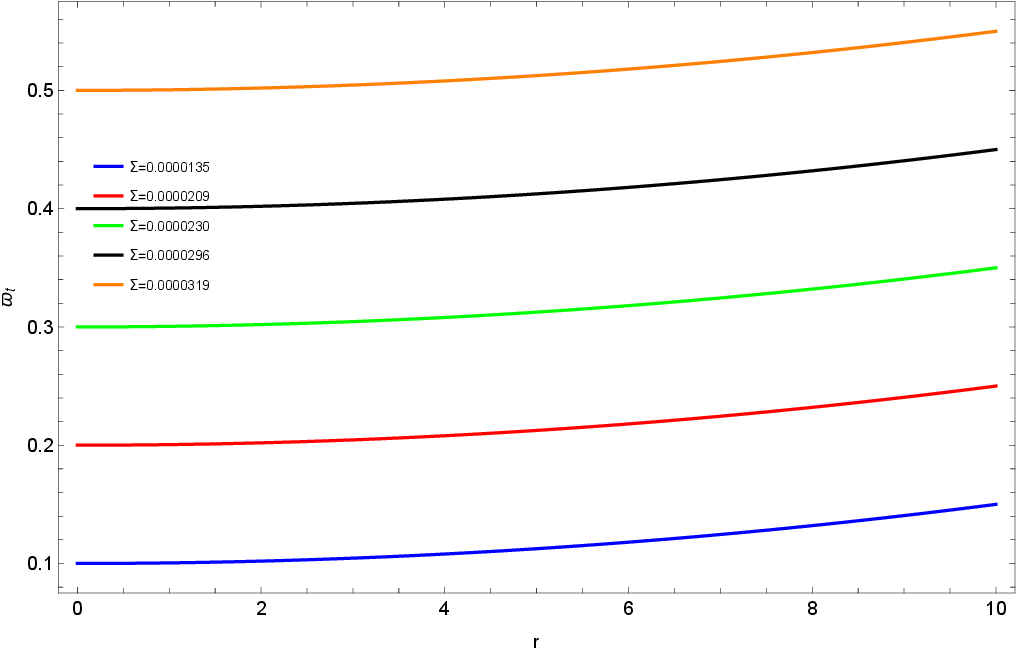}
\caption{Variation of $ \varpi_{t} $ for different parameter values}\label{9}
\end{figure}\\
\begin{figure}[ht!]
\centering
\includegraphics[scale=0.5]{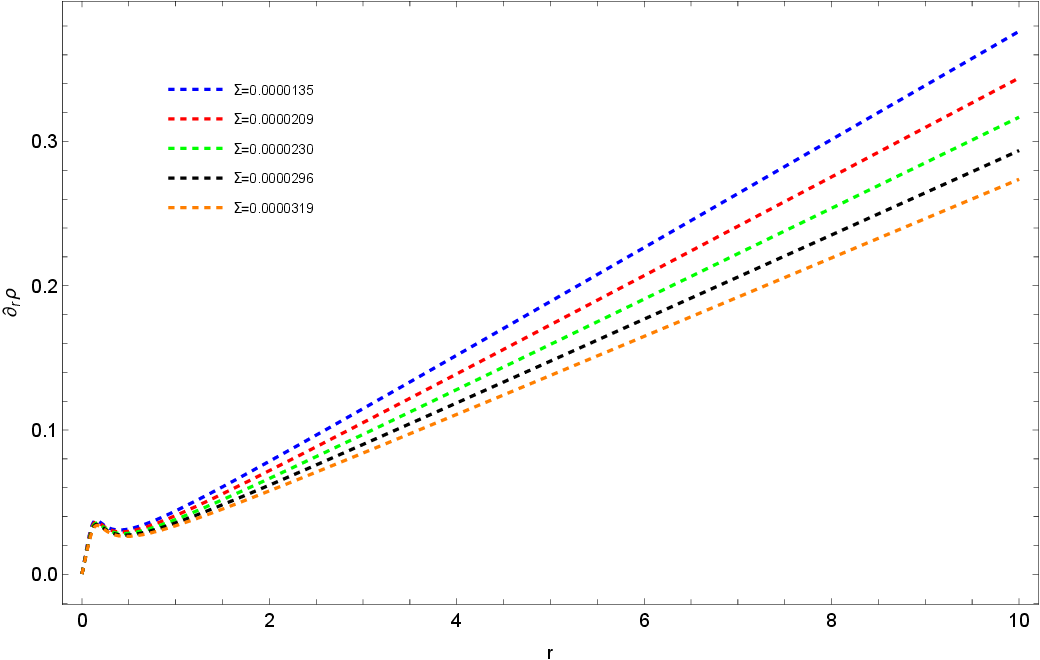}
\caption{Nature of radial gradient of energy density for different parameter values}\label{10}
\end{figure}\\
\begin{figure}[ht!]
\centering
\includegraphics[scale=0.5]{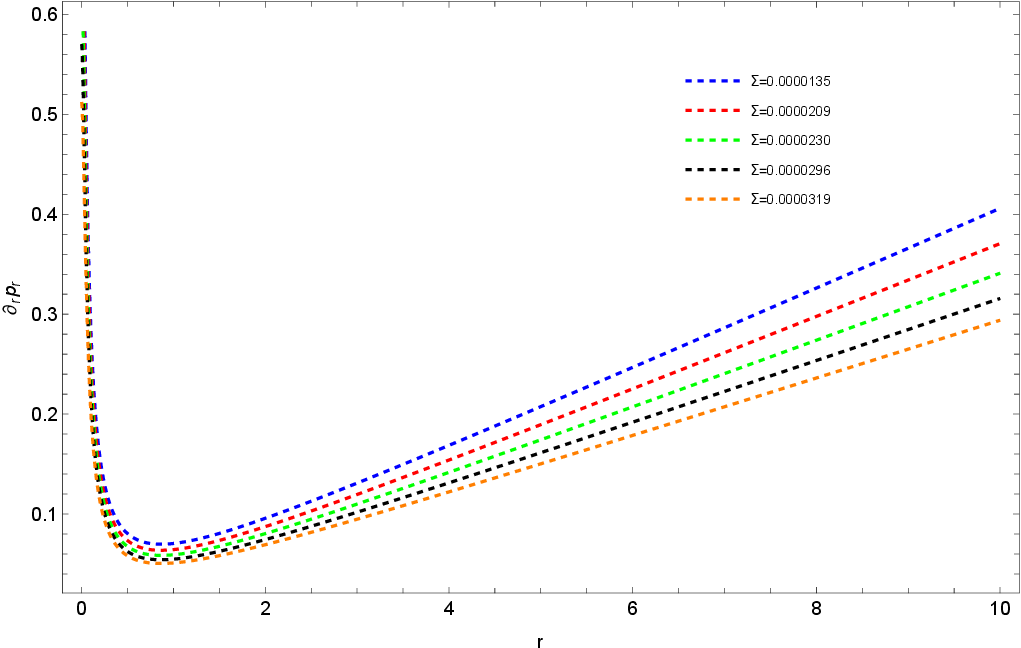}
\caption{Nature of radial gradient of $ p_{r} $ for different parameter values}\label{11}
\end{figure}\\
\begin{figure}[ht!]
\centering
\includegraphics[scale=0.5]{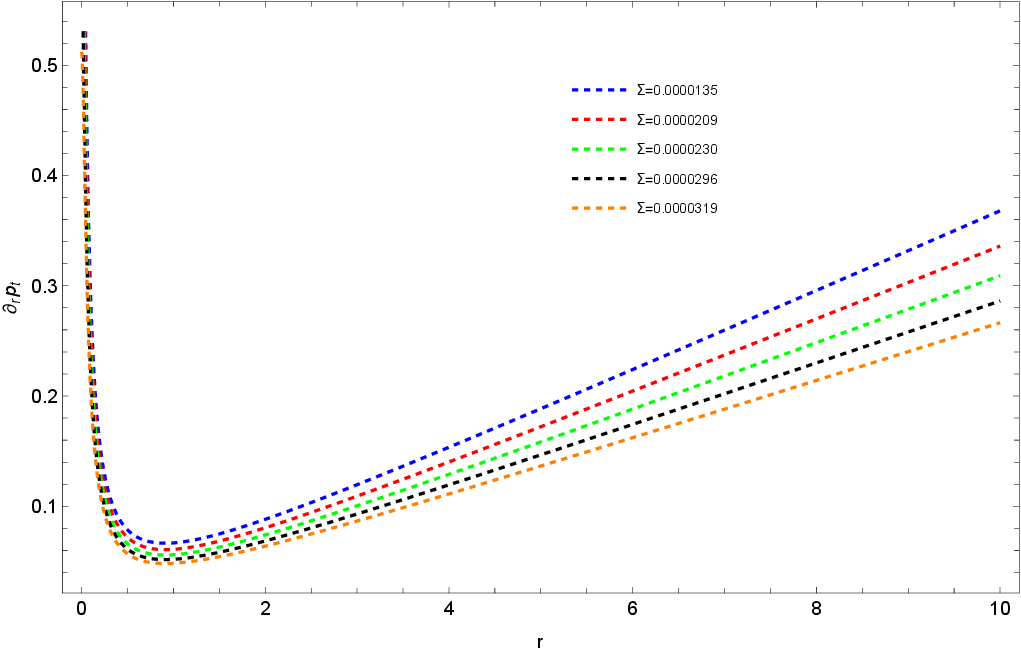}
\caption{Nature of radial gradient of $ p_{t} $ for different parameter values}\label{12}
\end{figure}\\

\subsection{TOV equilibrium}

\hspace{0.5cm}The modified TOV equilibrium equation below helps assess the viability of the matter distribution inside the star, which can be seen in the works of \cite{B29,B30,B31,B32}
\begin{equation}\label{21}
-\frac{dp_{r}}{dr} - \frac{j'(r)}{2}(\rho + p_{r}) + \frac{2}{r}(p_{t} - p_{r}) + F_{E} = 0. 
\end{equation}
The equation differs from the standard TOV equilibrium due to the additional force $ F_{E} $ which is crucial for accurately modeling the stable quark star configurations in mimetic gravity. Equilibrium is ensured by the extra force term in the modified equations. The above equation (\ref{21}) can be categorized as hydrodynamical, gravitational, anisotropic and extra force which are given as
\begin{eqnarray}
F_{H} = -\frac{dp_{r}}{dr}
\nonumber \\
F_{A} = \frac{2}{r}(p_{t} - p_{r})
\nonumber \\
F_{G} = - \frac{j'(r)}{2}(\rho + p_{r}).
\end{eqnarray}
Figures (\ref{13}-\ref{15}) display numerical solutions for the forces considering various parameter values. Without the extra force, solutions seem to be unstable, however introducing a small positive extra force renders them stable in this modified mimetic gravitational theory.
\begin{figure}[ht!]
\centering
\includegraphics[scale=0.5]{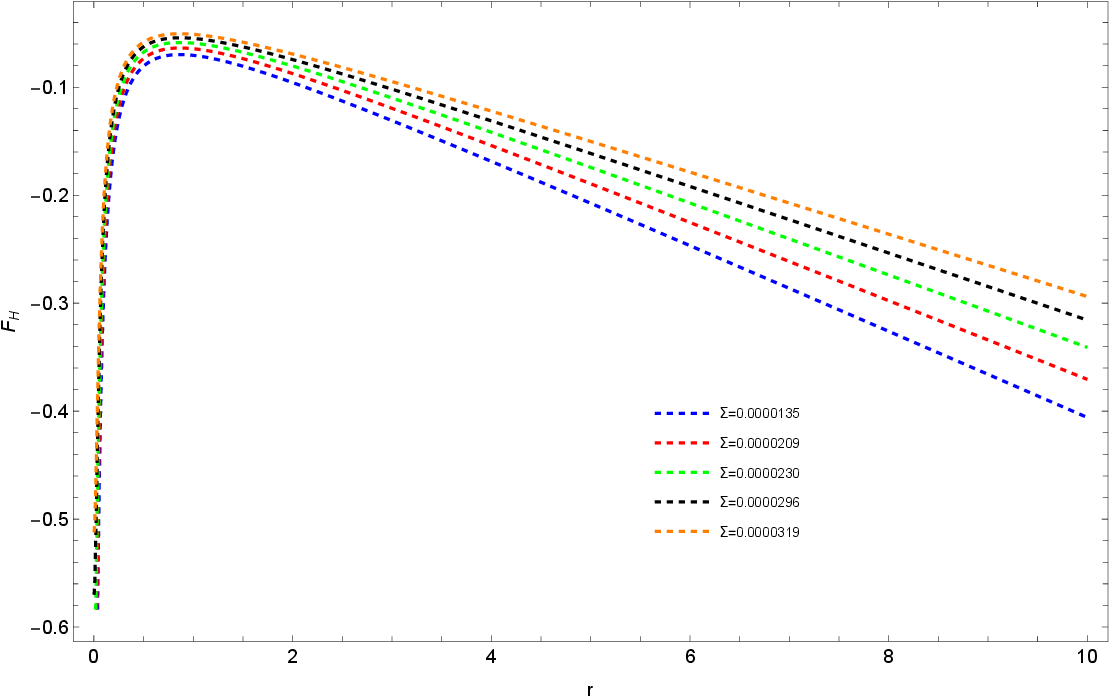}
\caption{Plot of $ F_{H} $ versus $ r $ for different parameter values}\label{13}
\end{figure}\\
\begin{figure}[ht!]
\centering
\includegraphics[scale=0.5]{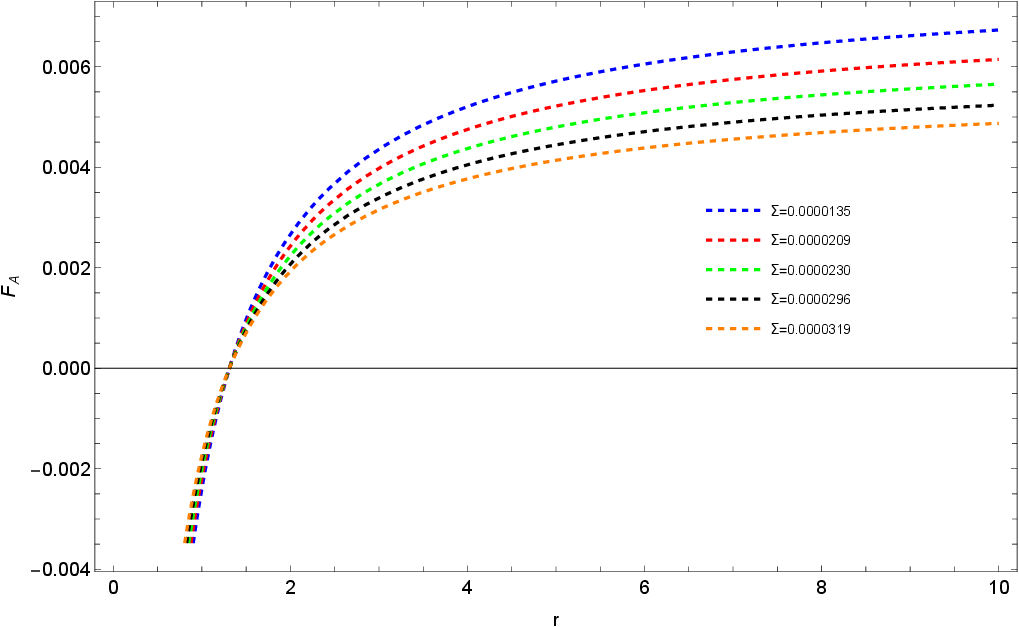}
\caption{Plot of $ F_{A} $ versus $ r $ for different parameter values}\label{14}
\end{figure}\\
\begin{figure}[ht!]
\centering
\includegraphics[scale=0.5]{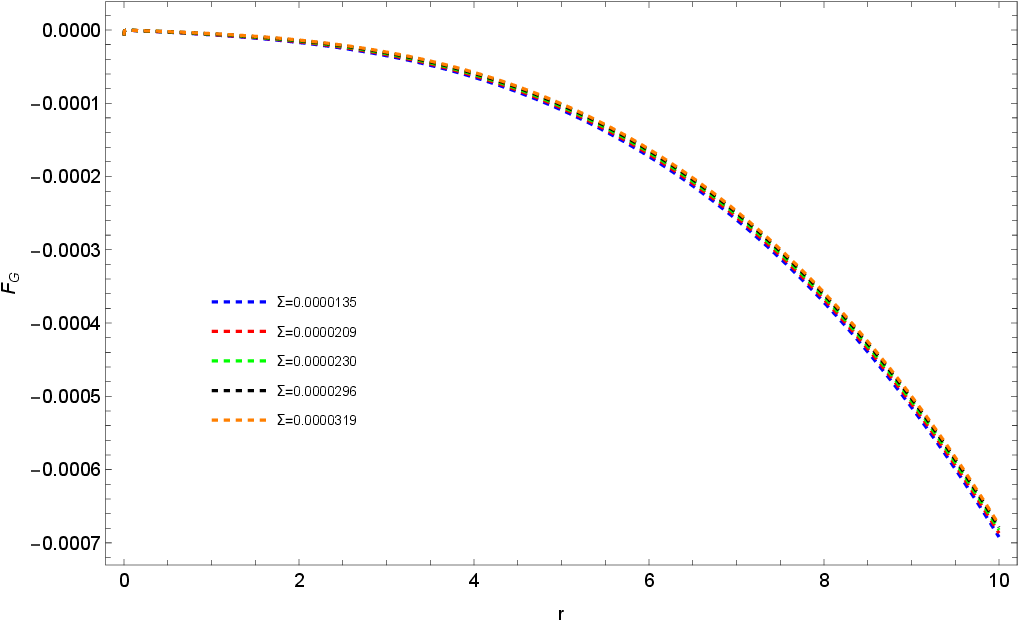}
\caption{Plot of $ F_{G} $ versus $ r $ for different parameter values}\label{15}
\end{figure}\\

\section{Stability analysis}\label{sec5}

\hspace{0.8cm}Next we discuss the stability of anisotropic quark stars within the framework of mimetic gravity theory. Three stability indicators are being considered here: adiabatic index, surface redshift and speed of sound analysis.\\

\subsection{Adiabatic index}

\hspace{0.5cm}The adiabatic index as derived in the seminal works of \cite{B33} helps in analysing compact relativistic solutions' stability under adiabatic perturbations. The ratio of specific heats, known as the adiabatic index ($ \Gamma $) is vital for compact star stability analysis. For anisotropic matter it has the following form as \cite{B34}
\begin{equation}\label{23}
\Gamma = (\frac{p_{r} + \rho}{p_{r}})\frac{dp_{r}}{d\rho}
\end{equation}
where
\begin{equation}\label{24}
\frac{dp_{r}}{d\rho} = \frac{dp_{r}}{dr}\frac{dr}{d\rho}.
\end{equation}
For compact objects to be dynamically stable, the adiabatic index must not exceed $ \frac{4}{3} $ \cite{D110}. $ \Gamma > \frac{4}{3} $ within the stellar interior confirms adiabatic stability of the solution. Figure (\ref{16}) presents the numerical solutions for $ \Gamma $. The strange quark stars here exhibit adiabatic stability throughout their interior for various parameter values. Larger parameter values might lead to instability near the stellar core. Thus our model appears to be consistent when viewed through the lens of the relativistic adiabatic index.\\
\begin{figure}[ht!]
\centering
\includegraphics[scale=0.5]{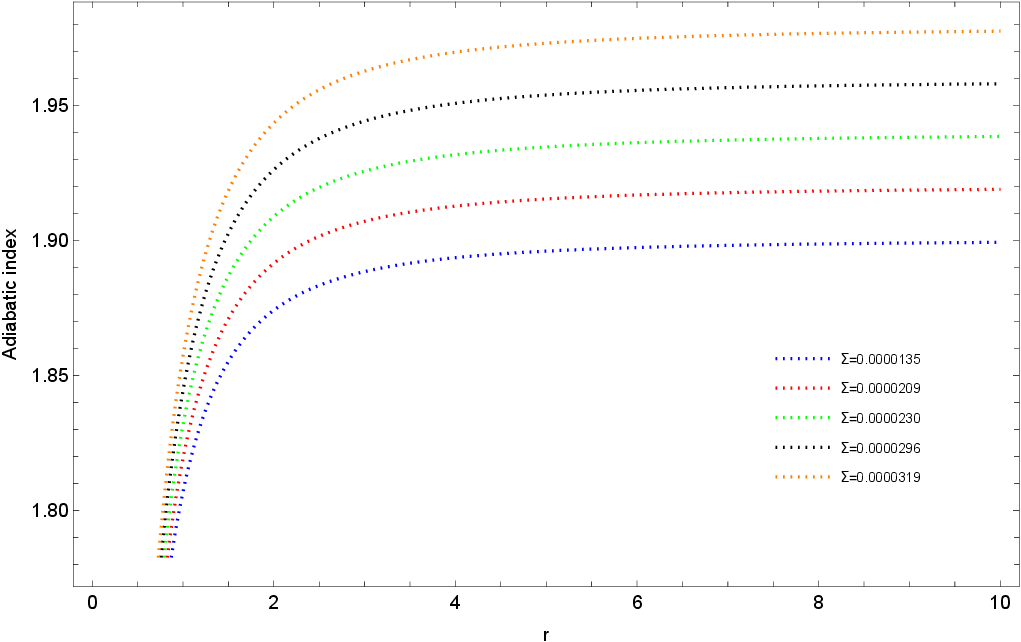}
\caption{Adiabatic index for different parameter values against $ r(km) $}\label{16}
\end{figure}\\

\subsection{Surface redshift}

\hspace{0.5cm}Surface redshift is the key to understand galaxy and cosmic characteristics. As electromagnetic radiation's wavelength stretches, this phenomenon takes place. Redshift indicates the change in wavelength from emission to reception. The surface redshift function is given by
\begin{equation}\label{25}
\verb"Z"_{s} = - 1 + \frac{1}{\sqrt{|g_{tt}|}}.
\end{equation}
Surface redshift limit for anisotropic matter is $ 2 $ or less. Figure (\ref{17}) displays solutions for different parameter settings for the redshift function. Clearly, from the figure we can see that for our model the surface redshift function is positive and stays well below the range of $ 2 $. Thus we can say that surface redshift requirements are met for our stellar configuration \cite{D,Q}.
\begin{figure}[ht!]
\centering
\includegraphics[scale=0.5]{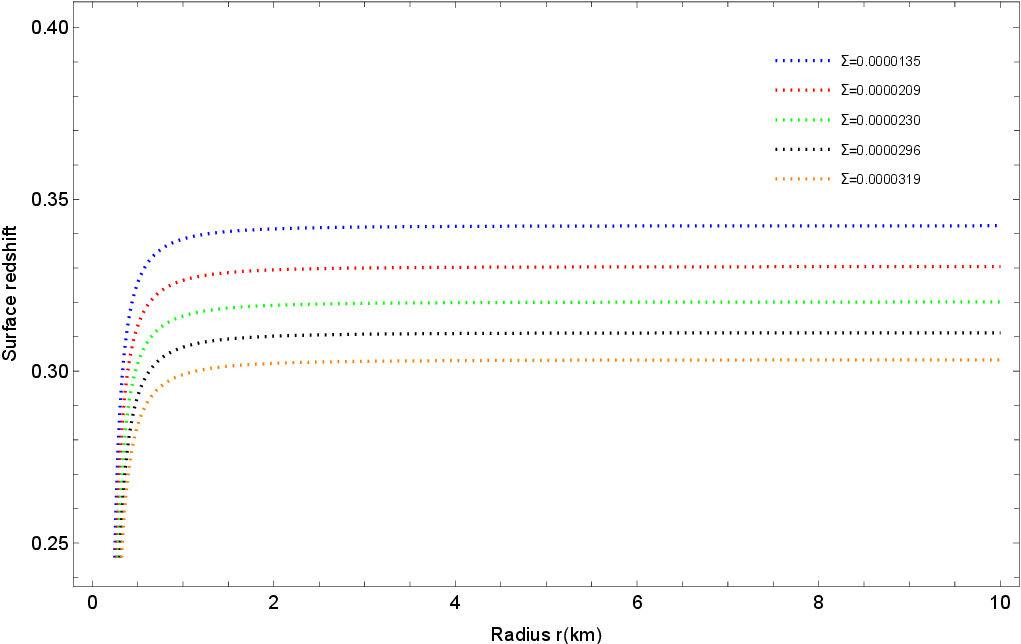}
\caption{Surface redshift function for different parameter values against $ r(km) $}\label{17}
\end{figure}\\ 

\subsection{Speed of sound analysis}

\hspace{0.5cm}Sound speed behavior has an influence on the quark stars stability. Casuality is verified via square of speed of sound measurement, i.e. $ \textsc{V}_{s}^{2} $ as
\begin{equation}\label{26}
\textsc{V}_{s}^{2} = \frac{dp_{r}}{d\rho} \leq c^{2} = 1. 
\end{equation}
Sound speed doesn't exceed light speed which we have demonstrated here. Our quark model satisfies casuality conditions, as is evident in figure (\ref{18}), where we can see that the parameter of sound of speed remains within the stable range of $ 0 $ and $ 1 $. Thus our mimetic gravity model solutions presented here appear physically plausible.
\begin{figure}[ht!]
\centering
\includegraphics[scale=0.5]{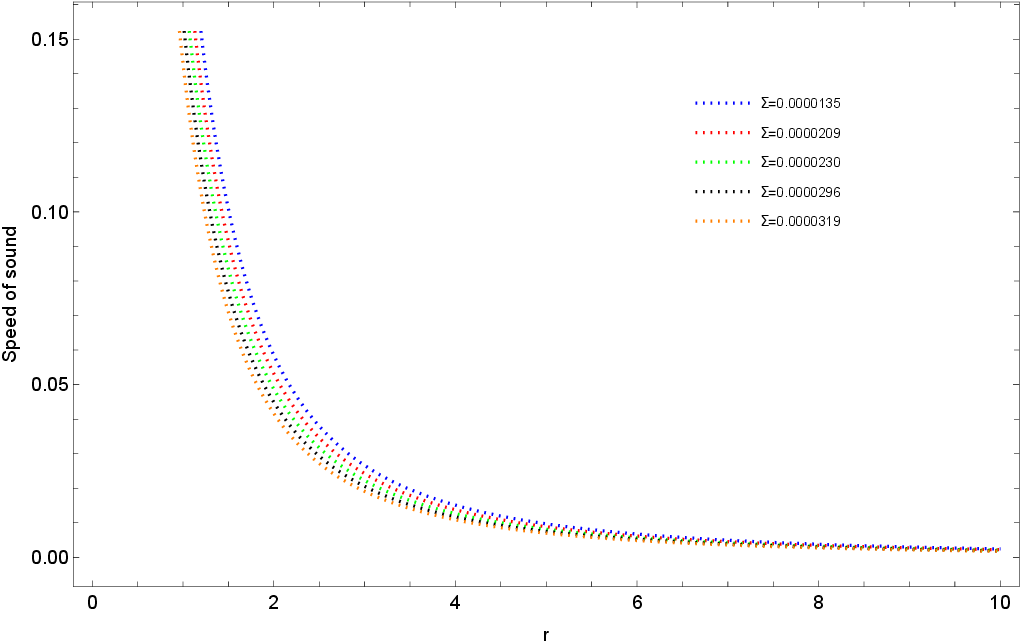}
\caption{Speed of sound analysis for different parameter values against $ r(km) $}\label{18}
\end{figure}\\ 

\section{Conclusion}\label{sec6}

\hspace{0.8cm}In this work, anisotropic spherically symmetric spacetime has been analyzed in mimetic gravity theory. Our investigation focuses on strange quark stars in this modified gravitational theory using the MIT bag model and the Buchdahl metric function. Here, metric coefficients are derived via matching the interior and exterior spacetimes. This analysis' core aspects are summarized as below.\\
\begin{itemize}
\item Our results show that the energy conditions namely NEC, DEC and SEC are satisfied within the stellar interior as demonstrated in figures (\ref{3}-\ref{7})
\item The EoS behaviors for the radial and tangential pressure are shown in the figures (\ref{8}) and (\ref{9}). Our findings show that the strange quark stars is physically viable for both the pressure types ( i.e. $ 0 \leq \varpi \leq 1 $).
\item Figures (\ref{10}-\ref{12}) demonstrate that for the metric tensor components' gradients, $ \partial_{r} \rho \ll 1 $, $ \partial_{r} p_{r} \ll 1 $, and $ \partial_{r} p_{t} \ll 1 $. This guarantees that the sound speed won't surpass the light speed.
\item We have then applied the TOV equilibrium condition to study the dynamical stability. We can see that a small extra force will ensure the physical validity in this framework. The findings are clearly depicted in figures (\ref{13}-\ref{15}).
\item Adiabatic index analysis constrains and stabilizes compact stellar objects. Mimetic strange stars in this case show overall adiabatic stability. The results are visualized in figure (\ref{16}).
\item The surface redshift in our model stays below $ 2 $, as is evident in figure (\ref{17}) which also supports the physical acceptability of the strange quark stars in mimetic theory.
\item For physical validity, the squared sound speed should lie between $ 0 $ and $ 1 $ to avoid tachyons. Figure (\ref{18}) confirms the physical stability through the squared speed of sound for mimetic strange quark stars.
\end{itemize}
Thus, the present analysis demonstrates that the Buchdahl metric potential provides a consistent and physically admissible framework for modeling mimetic strange stars. This work not only reinforces the viability of mimetic gravity in describing ultra-dense stellar configurations but also paves the way for further investigations employing alternative non-singular metric potentials. Such explorations hold promise for unveiling new insights into the behavior of matter under extreme conditions, thereby enriching our understanding of high-density astrophysical phenomena.\\

\end{document}